# Shadfa 0.1: The Iranian Movie Knowledge Graph and Graph-Embedding-Based Recommender System

Rayhane Pouyan
*Department of Computer engineering,
Center of Excellence on Soft Computing
and Intelligent Information Processing
Ferdowsi University of Mashhad*
Mashhad, Iran
r.pouyan@mail.um.ac.ir

Hadi Kalamati
*Department of Computer engineering,
Center of Excellence on Soft Computing
and Intelligent Information Processing
Ferdowsi University of Mashhad*
Mashhad, Iran
hkalamati@mail.um.ac.ir

Hannane Ebrahimian
*School of Electrical and Computer
Engineering
Faculty of Engineering
University of Tehran*
Tehran, Iran
h.ebrahimian98@ut.ac.ir

Mohammad Karrabi
*Department of Computer engineering,
Center of Excellence on Soft Computing
and Intelligent Information Processing
Ferdowsi University of Mashhad*
Mashhad, Iran
mohammad.karrabi@mail.um.ac.ir

Mohammad-R. Akbarzadeh-T
*Department of Computer engineering,
Center of Excellence on Soft Computing
and Intelligent Information Processing
Ferdowsi University of Mashhad*
Mashhad, Iran
akbarzadeh@ieee.org

*Abstract*— Movies are a great source of entertainment. However, the problem arises when one is trying to find the desired content within this vast amount of data which is significantly increasing every year. Recommender systems (RSs) can provide appropriate algorithms to solve this problem. The content-based technique has found popularity due to the lack of available user data in most cases. Content-based RSs are based on similarity of items demographic information; Term Frequency – Inverse Document Frequency (TF-IDF) and Knowledge Graph Embedding (KGE) are two approaches used to vectorize data to calculate these similarities. In this paper, we propose a weighted content-based movie RS by combining TF-IDF which is an appropriate approach for embedding textual data such as plot/description and KGE which is used to embed named entities such as director's name. The weights between features are determined using a Genetic algorithm. Additionally, Iranian movies dataset is created by scraping data from movie-related websites. This dataset and the structure of the FarsBase KG are used to create the MovieFarsBase KG which is a component in the implementation process of the proposed content-based RS. Using precision, recall, and F1 score metrics, this study shows that the proposed approach outperforms the conventional approach that uses TF-IDF for embedding all attributes.

## I. Introduction

Information Filtering is a kind of intelligent computing technique that mitigates this problem by providing the user with the most relevant information.

RSs are Information Filtering tools that direct users toward their interested, relevant products or services in a personalized way. Among various systems available to access information, the RS plays an important role in improving businesses and facilitating users' decision-making. Generally, the list of recommendations is created based on users' preferences and interests, items' properties, users' past interactions, and some additional information such as time and spatial data.

RS models, based on the types of input data, are mainly classified into 3 categories. Collaborative Filtering, Content-based Filtering, and Hybrid RS. Collaborative filtering and hybrid approaches rely on the user-item interactions which are not available in many applications. Content-based approaches are great models when there isn't enough interaction data and the recommendation is based on keywords and attributes assigned to items in a database.

In this paper, we propose a content-based movie RS based on movies' demographic information. We use publicly available features such as movie plot, genre, director's name, etc. to find similarity between movies and provide a recommendation list. In this regard, the textual data may need to be transformed to Vector Space Model (VSM). VSM is used in many search engines and, it performs well on tasks that required similarity among terms, phrases, and documents. TF-IDF is a way for converting textual data to VSM. It is a numerical statistic that is intended to reflect how important a word is to a document in a collection or corpus. Its limitation is that it does not take into account semantic similarities in named entities. The knowledge graph (KG) adds semantic information to data by considering the relationships between entities. In this paper we create the first Iranian Movie Knowledge Graph named MovieFarsBase which is used to add semantic information. We use a combination of TF-IDF and KGE to alleviate their limitations. We use TF-IDF to vectorize textual properties such as Movie plot/description, and Translating Embeddings (TransE) algorithm to model Multi-relational Data [1] such as Director's name, Producer's name, Actors' name, and Genre. The weights between these vectors are determined using a genetic algorithm. After creating these vectors for each movies, we calculated similarity between movies using the similarity metrics.

The structure of this paper is organized as follows: Section 2 reviews the literature of RSs; Section 3 introduces the collection of Iranian movies dataset; In section 4 we introduce MovieFarsBase Knowledge Graph; Section 5 addresses the movies similarity calculation; In section 6 we present the application of the Genetic algorithm to optimize the weights of attributes; and Section 7 provides experiments results and discusses and checks the effectiveness of various factors to evaluate our methodology; and finally in the section 8 is devoted to concluding discussion and suggestions for future works.

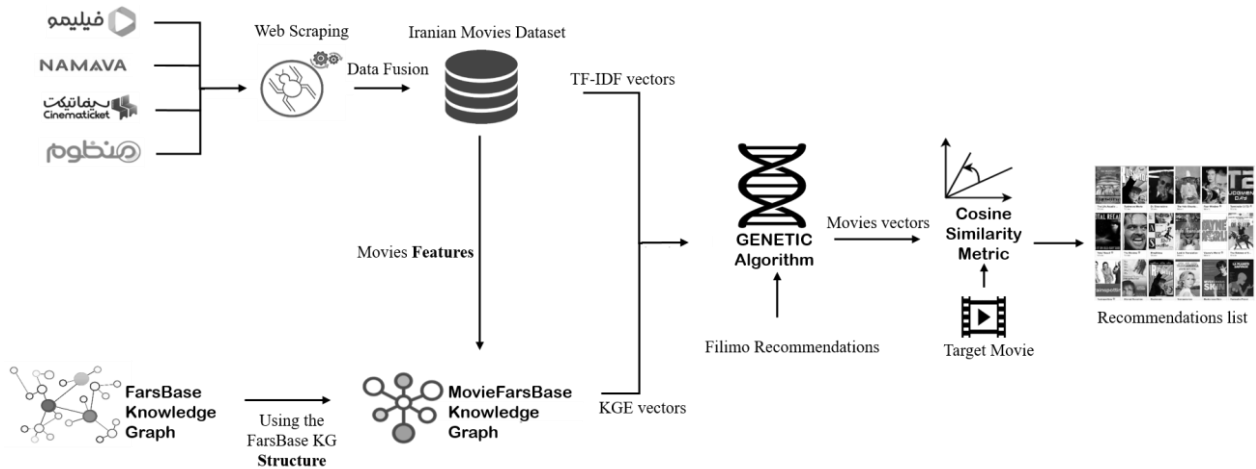

Fig 1- Block diagram of the proposed method. This research consists of 5 main stages including data collection, Dataset and KG construction, Data vectorization, optimization, and calculating similarities.

## II. RELATED WORKS

Considering the type of the available information, RSs are categorized into 3 main classes.

### A. Collaborative Filtering

Collaborative filtering systems recommend items based on similarity between users and/or items. The system recommends those items that are preferred by similar kinds of users. Manoj Kumar et al. in [2] presented a movie recommender system based on a collaborative filtering approach that makes use of the information provided by users, analyzes them, and then recommends the movies that are best suited to the user. The recommended movie list is sorted according to the ratings given to these movies by previous users and it uses K-means algorithm for this purpose. Lakshmi Tharun Ponnam et al. in [3] proposed an Item based collaborative filtering technique. They first examine the User item rating matrix and identify the relationships among various items, and then they use these relationships in order to compute the recommendations for the user.

### B. Content-based

Content-based filtering methods are based on a description of the item and a profile of the user's preferences. These methods are best suited to situations where there is known data on an item (name, location, description, etc.), but not on the user. In [4] Rujhan Singla et al. presented a movie recommendation framework (FLEX) following a content-based filtering approach. FLEX extends existing approaches like Doc2Vec and TF-IDF by using a hybrid of the two methods. They use publicly available features such as movie plots, ratings, countries of production and release year to find similarity between movies and generate a recommendation list.

### C. Hybrid approaches

Hybrid Recommender systems recommend items by combining two or more methods, including the content-based method, the collaborative filtering-based method, the demographic method, etc. A hybrid movie recommender system proposed by Bogdan Walek and Vladimir Fojtik in [5]. They presented a monolithic hybrid recommender system called Predictory, which combines a recommender module composed of a collaborative filtering system (using the SVD algorithm), a content-based system, and a fuzzy expert system. Geetha G et al. in [6] proposed a movie recommendation system that has the ability to recommend movies to a new user as well as the other existing users. It mines movie databases to collect all the important information, such as, popularity and attractiveness, which are required for recommendation. They use content-based and collaborative filtering and also hybrid filtering, which is a combination of the results of these two techniques, to construct a system that provides more precise recommendations concerning movies. In [7] a movie recommender system is built using the K-Means Clustering and K-Nearest Neighbor algorithms.

In this research, we focus on construction of the first Iranian Movie Knowledge Graph and development of a Graph-Embedding-Based Recommender System. We use TF-IDF to embed textual data and KG to model semantic information. Genetic algorithm is used to determine the optimal weights for features to get the best recommendations.

## III. PROPOSED METHOD

In this section, we describe the proposed content-based movie RS by combining TF-IDF and KGE. Fig 1 represents a block-diagram view of the entire logic of the proposed method. In the following, we describe the proposed method step by step.

### A. Iranian Movies Dataset Construction

*1) Data Collection:* Researchers often find it challenging to obtain data. The way to collect data has commonly been by observation, sample surveys, interviews or focus groups. In many cases, it requires time and money to gather information from the methods mentioned above. Nowadays, the Internet has become a significant source of information for many professionals and scientists. In this sense, to solve many problems from an academic point of view, it is easy, quick and cheap to extract information from the Internet. The development of computers has produced many useful techniques that can create massive databases. One technique is web scraping, which is used commonly to accumulate vast amounts of raw data. The term Web scraping refers to the process or technique of extracting information from various websites using specially coded software programs called Web Scraper or Crawler. Due to the lack of an Iranian movies dataset, creating a dataset is one of the essential steps of this research. For this purpose, we implement four crawlers to scrape data from the best Iranian movie websites named Filimo, Namava, CinemaTicket, and Manzoom. We utilize BeautifulSoup and Scrapy which are two popular python libraries to implement these crawlers. At last, we store attributes of all Iranian movies available on each website in a JSON file.

*2) Data Fusion:* Data fusion is the process of integrating information from multiple sources to produce specific, comprehensive, and unified data. As mentioned in the previous section, the collected data from each source store in a JSON file format. Because these sources may contain duplicate movies, the data fusion process involve identifying and deleting reduplicative data. The final data for a movie should cover all features available on all sources, and no information should be lost. We use the Levenshtein Distance [8] algorithm to identify duplicate movies. The Levenshtein distance is a string metric for measuring difference between two sequences. Mathematically, the Levenshtein distance between two strings $a, b$ (of $|a|$ and $|b|$ respectively) is given by $lev(a, b)$ where

$$lev(a,b) = \begin{cases} |a| & if\ |b| = 0 \\ |b| & if\ |a| = 0 \\ lev(tail(a), tail(b)) & if\ a[0] = b[0] \\ 1 + min \begin{cases} lev(tail(a), b) \\ lev(a, tail(b)) \\ lev(tail(a), tail(b)) \end{cases} & otherwise \end{cases} \quad (1)$$

Informally, the Levenshtein distance between two words is the minimum number of single-character edits (i.e. insertions, deletions or substitutions) required to change one word into the other.

We apply this algorithm on movie titles to identify duplicate movies between multiple sources. If there was same titles in multiple sources, only one of them would remain, and duplicate attribute values would be deleted. Final Iranian movies dataset contains 2739 unique movies. The attributes for each movies include Movie Title, Movie English Title, Storyline, Poster URL, Release year, Rates, Genre, Director, Producer, Actors, Duration, and Total sale.

### B. MovieFarsBase KG Creation

Knowledge graphs (KGs) organize data from multiple sources, capture information about entities of interest in a given domain or task (like people, places or events), and forge connections between them. The most common example is the Google knowledge graph, which is used in web search, or Amazon's product graph. Other knowledge graphs are openly available. These include DBpedia, Freebase, WordNet, FarsBase etc. In the following, more about FarsBase KG is explained.

*1) FarsBase Overview:* FarsBase is the first Persian knowledge graph that its data is extracted from Persian version of Wikipedia [9]. Similar to other English knowledge graphs, FarsBase is a collection of subject-relation-object triples($e_1, r, e_2$), where $e_1$ and $e_2$ are the entities (e.g., Hafez or Iran) and $r$ is a relation/predicate that connects two related entities like birthplace. A triple is also called a fact. FarsBase KG has 32,939,539 triples in 781 classes/domains.

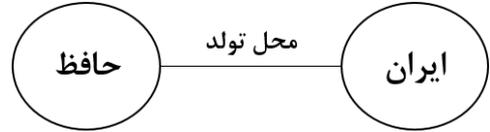

Fig 2-An example of a Tripl in FarsBase KG

*2) FarsBase enrichment by movie data:*

**Entity Matching**. It is a function that examines the existence of an entity in the FarsBase KG using SPARQL. Entities extracted from the created Iranian movies dataset, such as the names of people and the names of movies, are searched in the KG, and if they do not exist, a URI is defined for them.

**Inserting Data.** All the data in the dataset that did not already exist in the FarsBase KG were added to this KG as triples, observing the structure of the FarsBase KG.

*3) Detaching the movie subgraph (MovieFarsBase KG):* We already had the URI of the movies in the KG, we detach the size of the two hubs from the node of these movies in the KG. MovieFarsBase has 9,027,700 triples, 1,200,258 entities, and 6,320 relations.

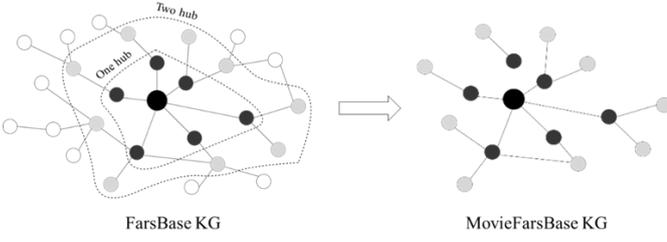

Fig 3-Detaching the movie subgraph

## C. Movies Similarity Calculation

To calculate the similarity between movies and generate a recommendation list we need to convert data from the textual format to Vector Space Model (VSM). VSM is developed by Gerard Salton in the SMART information retrieval system [10]. The idea behind VSM is to turn text documents from textual format to vectors where each document is a point in space that corresponds to a vector in vector space. Features are divided into two categories in terms of the technique used to vectorize them. Movie's plot/description and movie's title are textual features. We use TF-IDF to vectorize textual attributes. The other type of attributes such as director's name, producer's name, etc. are named entities. For this type of attributes, we need to use KGE models to transform property graphs to VSM. TransE is a simple and accurate KGE model that is used to embed MovieFarsBase KG. This model represents semantics by considering the relations of an entity in the KG. In the following, we examine TF-IDF, TransE, and Cosine similarity metric in more detail.

*1) TF-IDF:* Term frequency and inverse document frequency (TF-IDF) can recognize the important words or phrases in documents. It is the most common weighting method used to describe documents in the vector space model [22, 23]. If a word is infrequent but its number of occurrence is large in one or a few documents, it probably plays a key role to identify relevant recommendations. TF-IDF is calculated as a combination of the term frequency and inverse document frequency. TF is the number of times that a word $w$ occurs in the document $d$. Considering the length of documents, TF should be standardized.

$$TF = T/L \qquad (2)$$

Or

$$TF = T/T_i \qquad (3)$$

Where, $T$ is the term frequency, $L$ is the count of the unique words in document $d$, and $T_i$ denotes the frequency of the most frequent word in document $d$. IDF reveals how much information the word provides. It is calculated using $D$ and $D_i$, where $D$ denotes the number of all documents and $D_i$ is the number of the documents which include the word $w$.

$$IDF = \log \frac{D}{D_i + 1} \qquad (4)$$

Then,

$$TF - IDF = TF \times IDF \qquad (5)$$

In this scenario, document stands for movie and term denotes words in textual attributes.

*2) TransE:* Usually, we use a triple (head, relation, tail) to represent a knowledge. Here, head and tail are entities. TransE, or Translating Embeddings for Modeling Multi-relational Data, lets us embed the contents of a knowledge graph by assigning vectors to entities and relations. The basic idea of this model is making the sum of the head vector and relation vector as close as possible with the tail vector. It uses L1 or L2 norm to measure how much they are close.

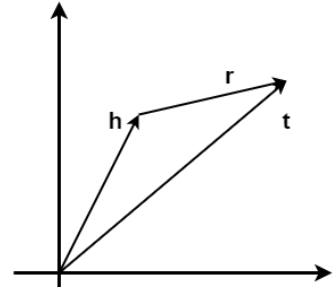

Fig 4- Vector representation of entities and relation

The loss function is the max-margin with negative sampling.

$$L(y, y') = \max(0, margin - y + y') \qquad (6)$$

$y$ is the score of a positive sample, and y' is the score of a negative sample. It is enough that the difference between the two scores is as large as margin (this value is usually 1). Because it uses distance to represent the score, so a minus are added to the equation, the loss function for knowledge representation is:

$$L(h, r, t) = \max(0, d_{pos} - d_{neg} + margin) \qquad (7)$$

And $d$ is:

$$d = |h + r - t| \qquad (8)$$

This is the $L1$ or $L2$ norm. As for how to get the negative sample is replacing the head entity or tail entity with a random entity in the triple.

*3) Cosine Similarity:* In Content-based Recommendation Systems, some specific similarity measures are used to find how equal two vectors of items (in our case vectors of movies) are in between them. In any kind of algorithm, the most common similarity measure is Cosine similarity. Mathematically, the cosine similarity is the cosine of the angle between two n-dimensional vectors in an n-dimensional space. It is the dot product of the two vectors divided by the product of the two vectors' lengths (or magnitudes). Cosine similarity is computed using the following formula:

$$similarity(A, B) = \frac{A.B}{|A| \times |B|} = \frac{\sum_{i=1}^{n} A_i \times B_i}{\sqrt{\sum_{i=1}^{n} A_i^2} \times \sqrt{\sum_{i=1}^{n} B_i^2}} \qquad (9)$$

Values range between -1 and 1, where -1 is perfectly dissimilar and 1 is perfectly similar. According to this metric,

the movies that are most similar to the target movie are included in the list of recommendations.

*D. Weight optimization using Genetic algorithm*

Attributes are not equally important for creating recommendation list. For this reason, we consider a weight for each attribute and use an optimization algorithm to find the optimal weights. When objective function and possibly constraint evaluations have a black-box formulation, derivative-free optimization approaches are commonly used. Genetic algorithm doesn't require any derivative information and also has very good parallel capabilities compared to the other counterparts. In this case, real number coding is used for Chromosome coding. To be specific, a Chromosome consisting of 5 genes, each of which determines the amount of each weight. The values of the weights are considered between 0 and 1 ($0 \leq w_i \leq 1$) because only the proportions of the weights are important (not their values). The Fitness function is based on a comparison of our algorithm's list of recommendations and the Filimo website.

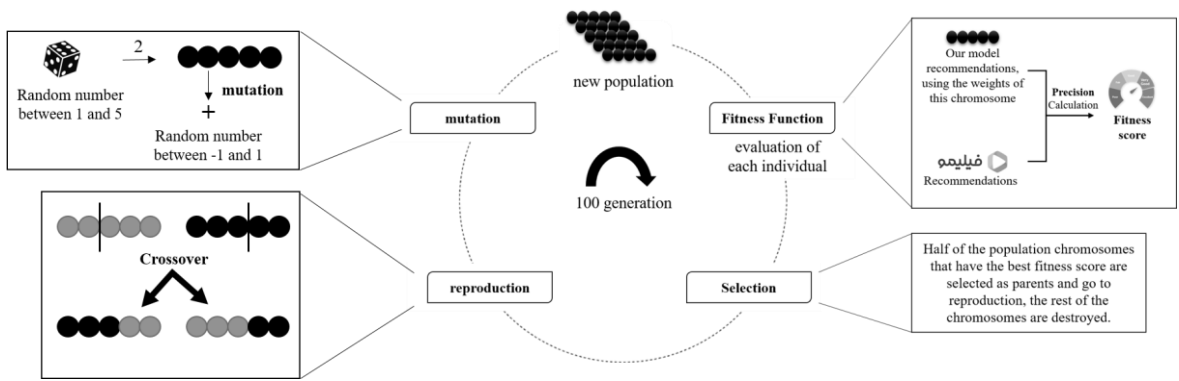

Fig 5- Genetic algorithm diagram

## IV. EXPERIMENT AND RESULTS

Contrary to a collaborative filtering system, where the predicted ratings can always be compared with actual ones, evaluating a content-based recommender system on the basis of statistical measures can be a daunting problem. This is mainly due to the absence of any 'ground truth' regarding its predictions. In order to evaluate the performance of our proposed algorithm, an experiment is conducted on the MovieFarsBase KG.

**Dataset.** MovieFarsBase KG is included 24,104 movies. Movie Title, Movie plot/description, Director's name, Producer's name, Actors' name and genre are used in this research.

**Experiment details.** The parameters in the genetic algorithm are set this way; the number of Chromosome in population is 8, the crossover rate and the mutation rate are 0.5 and 0.2, respectively, and up to 50 generations have been run. It should be noted that this algorithm is run in parallel on 8 cores.

**Evaluation.** A variety of evaluation criteria are used to obtain the subsequently mentioned results, each of which is described below.

*1) Coverage:* The coverage of a recommender system deals with the representation of the percentage of the training data that the recommender system was able to recommend. For a dataset containing $N$ items, out of which a total of $n$ items are recommended at some point, the coverage is given as:

$$Coverage = \frac{n}{N} * 100 \quad (10)$$

For a value of $N$ as 300 which is the size of the training data, a coverage of 95% was achieved using the proposed methodology.

*2) Precision@k:* We use Precision@k as one of the result metrics for evaluating the performance of our algorithm. Precision defines the ability of the system to propose content that is relevant for a user. It concerns a ratio of relevant recommendations with respect to all recommendations for the user. Precision can be calculated using the following:

$$Precision@k = \frac{Correctly\ recommended\ movies}{Total\ recommended\ movies} \quad (11)$$

*3) Recall:* Recall defines the ability of the system to provide the user with relevant content. It concerns the number of correct recommendations in a set of relevant recommendations. Recall can be calculated using the following:

$$Recall = \frac{Correctly\ recommended\ movies}{Relevant\ movies} \quad (12)$$

*4) F1 Score:* The F1-measure is then the harmonic mean between precision and recall:

$$F1\ Score = \frac{2 * Precision * Recall}{Precision + Recall} \quad (13)$$

Here, we treat the recommendation task as a classic binary classification problem. The recommendations made by our system are predictions and a relevant movie counts as a 'positive'. A 'true positive' here denotes movies which were relevant and were recommended by our system. We consider the movies that are recommended by Filimo and our system also recommend them as relevant movies. Because the number of Filimo movies recommended varies, k for each movie is calculated based on the number of Filimo movies Recommended.

*5) Human Evaluation:* As the identification of correct (relevant) and incorrect (irrelevant) items in the list of recommended items is highly subjective, and it is not accurate to consider the recommendations of Filimo as 'ground truth', we designed a questionnaire to collect data from users. Testing on the group of real users is one of the possibilities that has also been used in other research projects [5]. This questionnaire is designed for 56 movies and for each movie a combination of the results of the TF-IDF algorithm and the results of our proposed method is shown as suggestions. Users select similar movies for the specified movie. Using these data, Precision, Recall, and F1 Score criteria were calculated for both methods.

Table 1-Basic Statics about AHDMovieData Dataset

|  | **AHDMovieData Dataset** |
|---|---|
| **Version** | v1.0 |
| **Number of Users** | 33 |
| **Number of Movies** | 56 |
| **Number of Opinions** | 1717 |
| **Released** | January 24.2022 |
| **Format** | JSON |

**Results.** At first, we use only TF-IDF vectors for all attributes and calculate the Precision once without optimization algorithm and again with Genetic algorithm to optimize weights. Secondly, we use a combination of KG and TF-IDF methods for vectorization of attributes, we obtain the recommendations and calculate the Precision. Fig 6 shows the fitness value over the first 50 generations. As shown in **Error! Reference source not found.**, Genetic optimize weights and increase Precision, and the presence of KG vectors improve the results. This evaluation proves performance improvement using a Genetic algorithm.

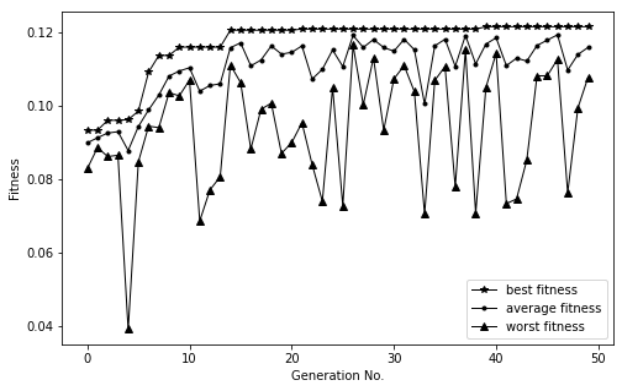

Fig 6-Fitness value in Genetic Algorithm.

Table 2-This table shows that Genetic algorithm improves the Precision in both aproaches

|  | **Precision@k** |
|---|---|
| **TF-IDF without Genetic** | 0.168 |
| **TF-IDF with Genetic** | 0.175 |
| **Proposed RS without Genetic** | 0.171 |
| **Proposed RS with Genetic** | 0.193 |

Human evaluation was performed with the participation of 33 people to compare the two methods. The results can be seen in the . The results show that our proposed method suggests more similar movies due to the use of KG.

Table 3-Human Evaluation Results for 56 movies

|  | **Precision@k** | **Recall** | **F1** |
|---|---|---|---|
| **TF-IDF** | 0.531 | 0.574 | 0.552 |
| **Proposed RS** | 0.672 | 0.644 | 0.658 |

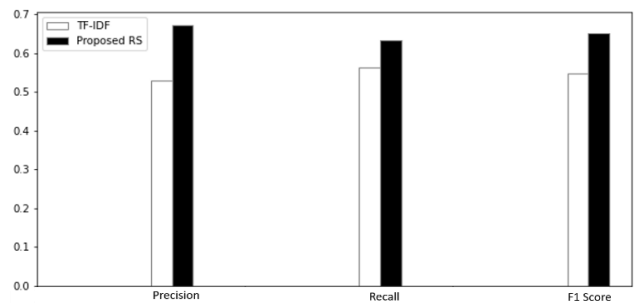

Fig 7- Precision, Recall, and F1 score assessments for Human Evaluation

## V. CONCLUSION AND FUTURE WORK

We develop a weighted content-based movie recommender system by combining TF-IDF and KGE. As far as we know, there is no similar recommender system or published method like what we have introduced here. Moreover, there is no dataset to use. Therefore, we first design multiple web crawlers to collect data and generate an Iranian movies dataset. The MovieFarsBase KG is created using this dataset and the FarsBase KG, to use it in implementing this RS. The proposed algorithm uses textual metadata of the movies like plot, cast, and genre to analyze them and recommend the most similar ones. Our system only needs a movie which the user is interested in to come up with suitable recommendations. We proved that the result obtained from the proposed RS outperforms the other mentioned approach (TF-IDF alone) using precision, recall, and F1 score metrics. Future work includes using deep learning KGE approaches for more accurate embedding KG and using newer methods like Bert instead of TF-IDF. Of course, using these new and highly complex techniques has more challenges, but it will help the quality of the RSs.

ACKNOWLEDGMENT

We thank our colleagues from AHD Company who provided insight and expertise that greatly assisted the research as well as providing financial support.

REFERENCES

[1]   A. Bordes, N. Usunier, A. Garcia-Duran, J. Weston, and O. Yakhnenko, "Translating Embeddings for Modeling Multi-relational Data," *Adv. Neural Inf. Process. Syst.*, vol. 26, 2013.

[2]   M. Kumar, D. K. Yadav, and V. K. Gupta, "A Movie Recommender System: MOVREC," *Int. J. Comput. Appl.*, vol. 124, no. 3, pp. 975–8887, 2015, Accessed: Nov. 08, 2021.

[Online]. Available: www.imdb.com.

[3] L. T. Ponnam, S. Deepak Punyasamudram, S. N. Nallagulla, and S. Yellamati, "Movie recommender system using item based collaborative filtering technique," *1st Int. Conf. Emerg. Trends Eng. Technol. Sci. ICETETS 2016 - Proc.*, Oct. 2016, doi: 10.1109/ICETETS.2016.7602983.

[4] R. Singla, S. Gupta, A. Gupta, and D. K. Vishwakarma, "FLEX: A content based movie recommender," *2020 Int. Conf. Emerg. Technol. INCET 2020*, pp. 8–11, 2020, doi: 10.1109/INCET49848.2020.9154163.

[5] B. Walek and V. Fojtik, "A hybrid recommender system for recommending relevant movies using an expert system," *Expert Syst. Appl.*, vol. 158, p. 113452, Nov. 2020, doi: 10.1016/J.ESWA.2020.113452.

[6] G. Geetha, M. Safa, C. Fancy, and D. Saranya, "A Hybrid Approach using Collaborative filtering and Content based Filtering for Recommender System," *J. Phys. Conf. Ser.*, vol. 1000, no. 1, p. 012101, Apr. 2018, doi: 10.1088/1742-6596/1000/1/012101.

[7] R. Ahuja, A. Solanki, and A. Nayyar, "Movie recommender system using k-means clustering and k-nearest neighbor," *Proc. 9th Int. Conf. Cloud Comput. Data Sci. Eng. Conflu. 2019*, pp. 263–268, 2019, doi: 10.1109/CONFLUENCE.2019.8776969.

[8] V. L.-S. physics doklady and U. 1966, "Binary codes capable of correcting deletions, insertions, and reversals," *nymity.ch*, 1965, [Online]. Available: https://nymity.ch/sybilhunting/pdf/Levenshtein1966a.pdf.

[9] M. Asgari-Bidhendi, A. Hadian, and B. Minaei-Bidgoli, "FarsBase: The Persian knowledge graph," *Semant. Web*, vol. 10, no. 6, pp. 1169–1196, Jan. 2019, doi: 10.3233/SW-190369.

[10] G. Salton, "The SMART retrieval system—experiments in automatic document processing," p. 156, 1971, [Online]. Available: https://dl.acm.org/doi/abs/10.5555/1102022.